\documentclass[journal]{IEEEtran}
\usepackage{graphicx}
\usepackage{stackengine}

\ifCLASSINFOpdf
\else
\fi
\usepackage{amsmath}
\usepackage{amsfonts} 
\usepackage{hyperref}
\usepackage{algorithm}
\usepackage{algpseudocode}
\ifCLASSOPTIONcompsoc
\else
\fi
\usepackage{caption}
\usepackage[labelformat=simple]{subcaption}
 
\usepackage{dsfont}
\usepackage[table,xcdraw]{xcolor}
\usepackage{textcase}
\makeatletter

\renewcommand{\section}{\@startsection{section}{1}{\z@}{-3.5ex \@plus -1ex \@minus -.2ex}{2.3ex \@plus.2ex}{\normalfont\normalsize\centering\bfseries\MakeUppercase}}
\renewcommand{\subsection}{\@startsection{subsection}{2}{\z@}{-3.25ex\@plus -1ex \@minus -.2ex}{1.5ex \@plus .2ex}{\normalfont\normalsize\MakeUppercase}}
\makeatother
\usepackage{paralist}

\begin{document}

\title{Bridging the Educational Divide: A Delay-Tolerant Networking Approach for Equitable Digital Learning in Rural Areas}

\author{Salah~Abdeljabar,~\IEEEmembership{Graduate~Student~Member,~IEEE,}
        and Mohamed-Slim~Alouini,~\IEEEmembership{Fellow,~IEEE}
\thanks{Salah Abdeljabar and Mohamed-Slim Alouini are with the Computer, Electrical and Mathematical Science and Engineering Division, King Abdullah University of Science and Technology (KAUST), Thuwal, Saudi Arabia (e-mail: salah.abdeljabar@kaust.edu.sa; slim.alouini@kaust.edu.sa).}
}

\markboth{}%
{Shell \MakeLowercase{\textit{et al.}}: Bare Demo of IEEEtran.cls for IEEE Journals}

\maketitle

\begin{abstract}
Access to quality education remains unequal, particularly in rural areas where Internet connectivity is limited or nonexistent. This paper introduces a framework for a digital learning platform that uses Delay Tolerant Networking (DTN) to extend educational opportunities to underserved communities. Unlike conventional models that rely on continuous Internet access, DTN offers an affordable and sustainable solution by leveraging existing transportation infrastructure. 
Beyond its technical contributions, the framework addresses ethical imperatives by promoting educational equity and digital inclusion.
We present a prototype tested on a university campus, demonstrating the feasibility of DTN for educational delivery. 
By addressing the digital divide, this framework aligns with global goals of inclusive education and sustainable development.
\end{abstract}

\begin{IEEEkeywords}
Delay Tolerant Networking (DTN), e-learning, Educational Digital Divide.
\end{IEEEkeywords}

\IEEEpeerreviewmaketitle
\section{Introduction}

\subsection{Motivation}
\par
In today's digital age, access to the Internet has become a fundamental requirement for economic and social growth. In fact, Internet connection is of paramount importance, so the United Nations (UN) declared in their common agenda that universal access to the Internet by 2030 is a basic human right \cite{Secretary-General_2021}. Nevertheless, according to the latest International Telecommunication Union (ITU) reports, around one-third of the world's population, some 2.9 billion people, remain with no access to the Internet, most of whom live in developing countries and rural areas \cite{ITU_Hub_2023}. 
This disparity creates what is commonly referred to as the digital divide, which not only limits access to information but also reinforces pre-existing inequalities related to income, geography, gender, and disability \cite{UNDigitalDivide2025, UNICEFGender2024}. 

\par
In recent years, a significant transformation has reshaped our approach to education. The traditional classroom model, where students solely rely on teachers' presentations, is giving way to a new paradigm. Extensive research has already highlighted the positive impact of information and communication technologies (ICTs) on education, including improved access to abundant online resources and digital learning content \cite{change2020delivering}.
Online learning platforms and educational resources accessible via the Internet offer rural communities the opportunity to access quality education and skill development programs that might not be readily available locally. This empowers individuals and equips them with the necessary knowledge and skills for socio-economic advancement. 
One of the key components to accessing such digital content is to have a robust ICT infrastructure, which is lacking in many rural and remote areas.

\subsection{Digital Learning for rural areas}
\par
Traditional learning methods include the establishment of physical schools or colleges along with the recruitment of faculty and staff. However, creating such institutions and hiring full-time educators requires substantial resources in rural areas. Moreover, professionally trained teachers often prefer urban areas for work and residence, leaving rural populations without access to quality educators and, consequently, quality education \cite{hussain2013learning}.
To address this disparity, online learning platforms can provide high-quality education with up-to-date learning materials \cite{change2020delivering}. 
Some of the components provided by online resources are access to visual or readable materials, such as Wikipedia, Khan Academy videos, and other educational materials.
Numerous e-learning platforms are available online, requiring continuous Internet connectivity. Alternatively, e-learning can be locally hosted in rural areas where uninterrupted Internet access cannot always be guaranteed. However, to ensure that the platform remains up-to-date with the latest resources and allows users to collaborate beyond geographical reach, periodic Internet connectivity is at least necessary for the platform for content updates.

\subsection{ICT requirements for digital learning}
\par
Improving access to education in rural areas through the use of ICT requires a concerted effort to overcome isolation and connectivity challenges. 
As evidenced in \cite{madimabe2021investigating}, on rural secondary education in South Africa, insufficient equipment and limited access to connectivity hinder the widespread adoption of digital learning processes. Additionally, the high cost of ICT infrastructure presents another barrier to implementing effective digital learning initiatives in rural regions. Similar considerations and challenges apply to both higher and primary education and are highlighted in \cite{gama2022electronic}. 
The impact of digital learning on rural communities, as highlighted in the study on \cite{anand2012learning}, goes beyond knowledge acquisition and technology skills. It also plays a crucial role in fostering social and intellectual development, ultimately contributing to the overall well-being and progress of rural populations.

\subsection{Access to the Internet in schools}
It is imperative for schools to have access to the Internet, as digital skills have become essential for young people to enter the labour market~\cite{ITU_Hub_2023}. 
The significance of Internet access in schools has been stressed during the COVID-19 pandemic, where schools without Internet connectivity faced challenges transitioning to online learning when required to close physically. 
However, data collected by UNESCO reveals that there are still many schools around the world that lack access to the Internet~\cite{ITU_Hub_2023}. Globally, approximately 40 per cent of primary and 66 per cent of secondary schools had access to the Internet. In the least developed countries (LDCs), these numbers were even lower, with 28 per cent of primary schools and 35 per cent of secondary schools having Internet access. While there have been improvements over the past years, access to the Internet in schools is still lacking in many regions. One such example to address the digital divide in education is a joint effort by ITU and UNICEF with the Giga initiative~\cite{Giga}, with the mission to connect schools worldwide to the Internet.
Currently, Giga has mapped around 1 million schools in mostly lower-income countries out of an estimated 6 million schools worldwide. However, data from UNESCO reveals that approximately 43 per cent of these mapped schools still lack Internet connectivity, indicating the need for further efforts to bridge the digital gap.

\subsection{Backhaul technologies in rural areas}
\par
Bringing traditional cellular links to rural areas to extend network coverage can often be cost-prohibitive. Compared to urban locations, remote and rural connectivity faces barriers in large geographical distances, complex terrain, and obstacles such as forests, mountains, or lakes. In addition, the low population density of rural locations and socio-economic factors potentially imply that the average revenue per mobile cell site will be much lower than in urban environments. Hence, in long backhaul stretches traversing non-populated areas, appropriate towers and repeaters need to be built to reach rural areas, which requires substantial investment that is lacking in many rural areas~\cite{yaacoub2020key}.  
\par
Another appealing solution is to connect rural areas with satellite networks. Various satellite connection providers offer telecommunication services and Internet access worldwide.  
Particularly, many companies have grand ambitions for their low Earth orbit (LEO) constellations, including SpaceX's Starlink and OneWeb, which have promising potential to provide scalable broadband connectivity for remote areas~\cite{osoro2021techno}.
Despite the effectiveness of satellite networks to provide extensive coverage and reasonable speeds, the reality remains that they involve high costs in installation and maintenance. Even the more decently priced options can still be quite expensive, with annual billing being common.
\par
On the other hand, delay tolerant networking (DTN) framework was proposed as a backhaul connectivity solution for rural areas~\cite{perumal2022comprehensive}.
In fact, when connectivity is absent or limited, DTN can rely on data mules like cars, trucks, and buses to carry data from remote rural areas to urban centres where connectivity to the Internet is provided. Thus, data can be stored in local nodes in remote areas until a DTN data mule collects it and carries it to the urban area. This idea of opportunistic communication needs little infrastructure investment yet can serve as a viable backhaul connection, especially for applications that can tolerate connection delays and disruptions.
While many technical efforts have focused on improving coverage through cellular or satellite solutions, these often overlook the social realities and constraints of rural communities. The DTN-based approach explored here prioritizes ethical principles of fairness, sustainability, and inclusivity by building upon existing infrastructure and minimizing costs, aligning with global efforts for equitable access to the digital world \cite{UNDigitalDivide2025}.

\subsection{Overview of this article}
This article explores the DTN framework's potential as a viable opportunistic backhaul solution to facilitate digital learning in rural areas; essentially, we try to address the digital divide in education via DTN.
The proposed framework empowers rural learners with access to a reliable digital learning platform, thereby enhancing their engagement with the world of ICT. 
To achieve these aims, we provide a brief overview of the DTN framework and its operation. Subsequently, we highlight the proposed DTN-Enabled digital learning platform solution and its properties. We then detail the prototype we designed and tested on our university campus before concluding the article.

\section{Background: Delay Tolerant Networking}
\begin{figure}[h!] 
  \centering
  \includegraphics[width=1\linewidth]{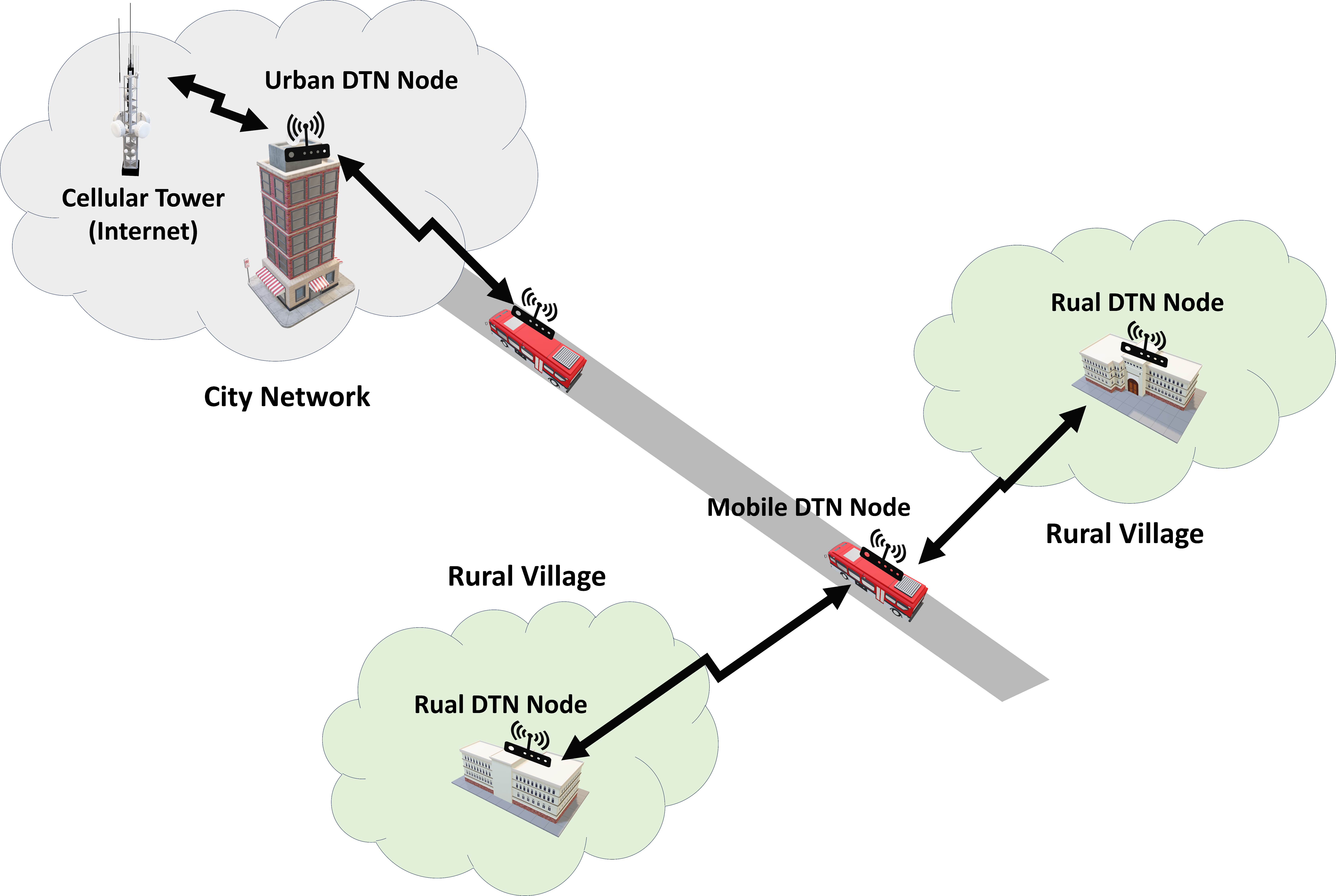}
  \caption{
  DTN connectivity architecture for rural areas. The system connects remote schools through mobile data transport via existing transportation infrastructure, creating an affordable and sustainable connectivity solution.
  }
  \label{fig:DTN_ProposedSystem}
\end{figure}

Traditional Internet rely on continuous, reliable connections between devices, allowing data to travel quickly and directly from sender to receiver. This setup works well in urban areas with good infrastructure but faces significant challenges in rural or remote regions where such connections are often unavailable or unreliable \cite{rodrigues2020advances}. 
DTN offers a different approach suited to these challenging environments. Instead of expecting constant connections, DTNs are designed to handle situations where devices connect intermittently and unpredictably. In these networks, data is stored temporarily on mobile devices, carried around as the devices move, and forwarded when the devices come into contact with others. This store-and-forward method allows information to eventually reach its destination, even if the connection is sporadic or delayed. 
One of the key advantages of DTN is that it requires very little infrastructure, making it an attractive solution for resource-constrained regions. 
Two key characteristics make DTN particularly valuable for rural development:
First, the system requires minimal infrastructure investment. As shown in Figure~\ref{fig:DTN_ProposedSystem}, it builds upon existing transportation networks and simple wireless technologies. Rural schools need only a basic DTN terminal with appropriate wireless capability to participate in the network. The data mules, which are vehicles equipped with DTN terminals, require no special routing or scheduling beyond their normal operations.
Second, the technology enables meaningful connectivity despite temporal delays. While real-time applications prove challenging, many essential services, including educational content delivery, health record synchronization, and market information sharing, function effectively with periodic updates.

\subsection{DTN for Rural and Underserved Areas}
Despite its origins in space communication, DTN has proven to be valuable for connecting rural and underserved communities that lack reliable traditional communication infrastructure. 
In many parts of the world, remote villages and isolated regions face significant barriers to accessing information, education, healthcare, and economic opportunities due to limited or unreliable connectivity. 
DTN offers a practical and sustainable approach to overcoming these challenges by leveraging existing transportation systems.
Researchers have explored various ways to implement DTN in rural settings. 
For example, in \cite{rahman2013delay}, the authors propose using public transport buses as mobile communication nodes to establish connections between remote villages in Bangladesh. This approach transforms everyday transportation into a means of digital connectivity, enabling communities to exchange vital information without the need for expensive infrastructure investments.
Similarly, in \cite{grasic2014revisiting}, DTN deployments in mountainous villages in Sweden utilize helicopters as mobile relay points, demonstrating how technology can adapt to challenging environments, ensuring that even the most isolated communities can participate in the digital age. 
Further studies, such as \cite{ntareme2011delay}, showcase the deployment of DTN over Android devices, turning everyday smartphones into portable communication nodes. This approach empowers individuals in remote areas to maintain contact with their communities and access information during times of crisis or infrastructural failure.

\section{Digital Learning Platform for Rural Schools Through DTN}
\par
\begin{figure}[h!]
    \centering
    \includegraphics[width=1\linewidth]{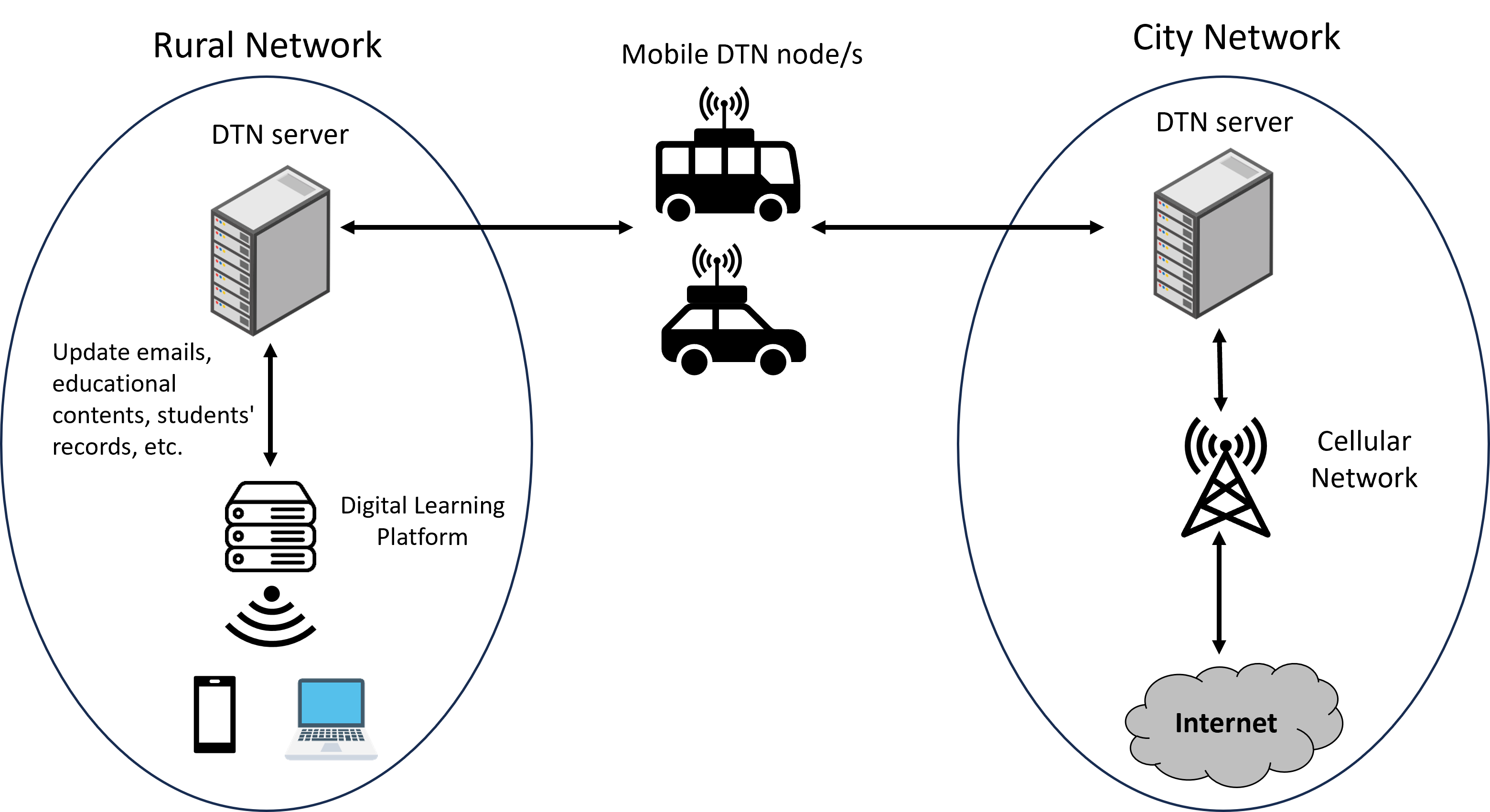}
    \caption{How the rural education network operates: (1) Schools host local digital learning platforms; (2) Mobile nodes on vehicles exchange data when in range; (3) Urban gateways connect to the Internet. This creates an affordable, sustainable solution for remote areas.
    }
    \label{fig:DTN_SystemArhitecture}
\end{figure}

In rural communities where Internet access is unreliable or nonexistent, students and teachers face significant barriers to digital education. Our proposed solution combines a digital learning platform with DTN, designed to provide educational resources even in areas with poor connectivity. This approach addresses not just technical challenges, but also the broader social inequities in education access. 
While traditional online platforms require constant Internet access, our system operates differently. It stores educational resources locally and synchronizes with the Internet only when a connection becomes available through mobile DTN nodes. These nodes, attached to vehicles like buses or trucks, act as digital couriers, transporting data between rural schools and urban areas with Internet access.
The system architecture of the proposed framework is shown in Figs.~\ref{fig:DTN_ProposedSystem} and \ref{fig:DTN_SystemArhitecture}. Rural networks consist of a DTN node and a digital learning platform. Users in rural areas can access the platform using a Wi-Fi connection, making it more accessible to a large segment of users. The rural network is opportunistically connected to the Internet via mobile DTN nodes, which will carry the data between the rural and city networks.

\subsection{DTN Communication Nodes}
The framework's components work together to create educational opportunities where they are needed most. In this part, we highlight the DTN communication nodes that facilitate connecting rural areas to urban areas with Internet connectivity.
\begin{inparaenum}[1)]
\item Rural DTN Node: Each rural school maintains a local server with educational materials. When a bus or other vehicle with a DTN node comes within range, the system automatically exchanges data, such as uploading student work and downloading new resources. These nodes are highlighted in Fig.~\ref{fig:DTN_ProposedSystem}.
\item Mobile DTN Node: Vehicles that regularly travel between rural and urban areas (like school buses or delivery trucks) carry small computing devices. These automatically collect and deliver educational content as they move between areas with and without Internet access. In Fig.~\ref{fig:DTN_ProposedSystem} we show an example of a bus carrying a DTN node with a wireless interface. Once the node is within the range of the remote school, it establishes a connection with the rural nodes.
\item Urban DTN Node: Located in urban areas, this node functions as a gateway between the DTN network and the Internet. It processes DTN traffic from remote locations, forwards data through the Internet, and supports updates from remote administrative servers to the mobile DTN node.  
\end{inparaenum}
These nodes establish wireless connections opportunistically based on the adopted technology. One appealing solution is to use TV white spaces (TVWS), which are the unused TV broadcast frequencies predominantly available in rural regions. This technology emerged as a sustainable technology covering large areas with challenging terrain while preserving energy~\cite{yaacoub2020key}.

\subsection{Empowering Learners Through Adaptive Features}
To create an enriching educational experience that empowers learners to advance their careers, the digital learning platform must incorporate essential features that guarantee high-quality content for its users. Leveraging the opportunistic Internet connection facilitated by DTN, the platform can offer specific features to enhance its functionality. These features include: 
\begin{inparaenum}[i)]
\item Access to Massive Open Online Courses: Massive Open Online Courses (MOOCs) offer free, flexible learning opportunities to anyone seeking to acquire new skills and advance their careers. While MOOCs typically require continuous Internet access, some courses can be downloaded for offline use \cite{oyo2014massive}. DTN enables users to request access to courses, which can then be transferred to the local platform, ensuring learning materials are available even during intermittent connectivity. 
\item Email: Seamless communication and interaction between students, teachers, and others are promoted through the option for users to access their emails~\cite{lindgren2007experiences}. This feature facilitates efficient information dissemination and enables schools to maintain strong connections with their educational community. Email retrieval, which doesn't require constant Internet access, can be handled by DTN.
\item Web-cashing: Leveraging DTN's opportunistic connectivity, the platform may employ web-caching to allow users to access preselected websites even without a continuous Internet connection. Web pages are retrieved from an Internet-connected server, transferred via DTN to the platform server, and can include a variety of resources such as weather updates, news articles, and educational content~\cite{naslund2013developing}.
\end{inparaenum}

\section{Bridging the Education Gap: A Real-World Test}

We developed and tested a practical solution to bring digital education to remote areas with poor or non-existent Internet access. This prototype demonstrates how technology can adapt to challenging environments while creating meaningful educational opportunities.
The framework was designed with three key principles in mind: First, \textit{accessibility}, where it works without a constant Internet connection. Second, \textit{affordability}, where we use low-cost components and existing infrastructure. Third, \textit{adaptability}, where the technology is scalable and can grow with local communities' needs. While technical performance metrics like data transfer rates are important (analyzed in~\cite{DTN_Abdeljabar2025}), our focus here is on how this technology can create a real educational value for underserved communities.

\subsection{How the System Helps Learners}
A typical digital learning platform provides users with access to various educational content. 
Our prototype focuses on providing text-based educational content to users in rural areas, where the learning platform is provided as a web application so anyone connected to the local network can access it. 
The platform stores educational materials on a school server, so they are available anytime without Internet. Also, it allows the students to request any topic they would like to learn about from Wikipedia, and then the request is sent via DTN to the urban area with Internet access. The request is processed there, and the content is sent back to the rural node via DTN. Additionally, the application allows educators to update existing material locally and publish new material to make it available to students with access to the digital learning platform.
This approach mirrors how rural communities have historically shared information through mobile libraries or travelling teachers, but with digital efficiency and rich learning experience, adapting to the limited resources.

\subsection{Building Blocks for Change}
\begin{figure}[h]
    \centering
    \includegraphics[width=1\linewidth]{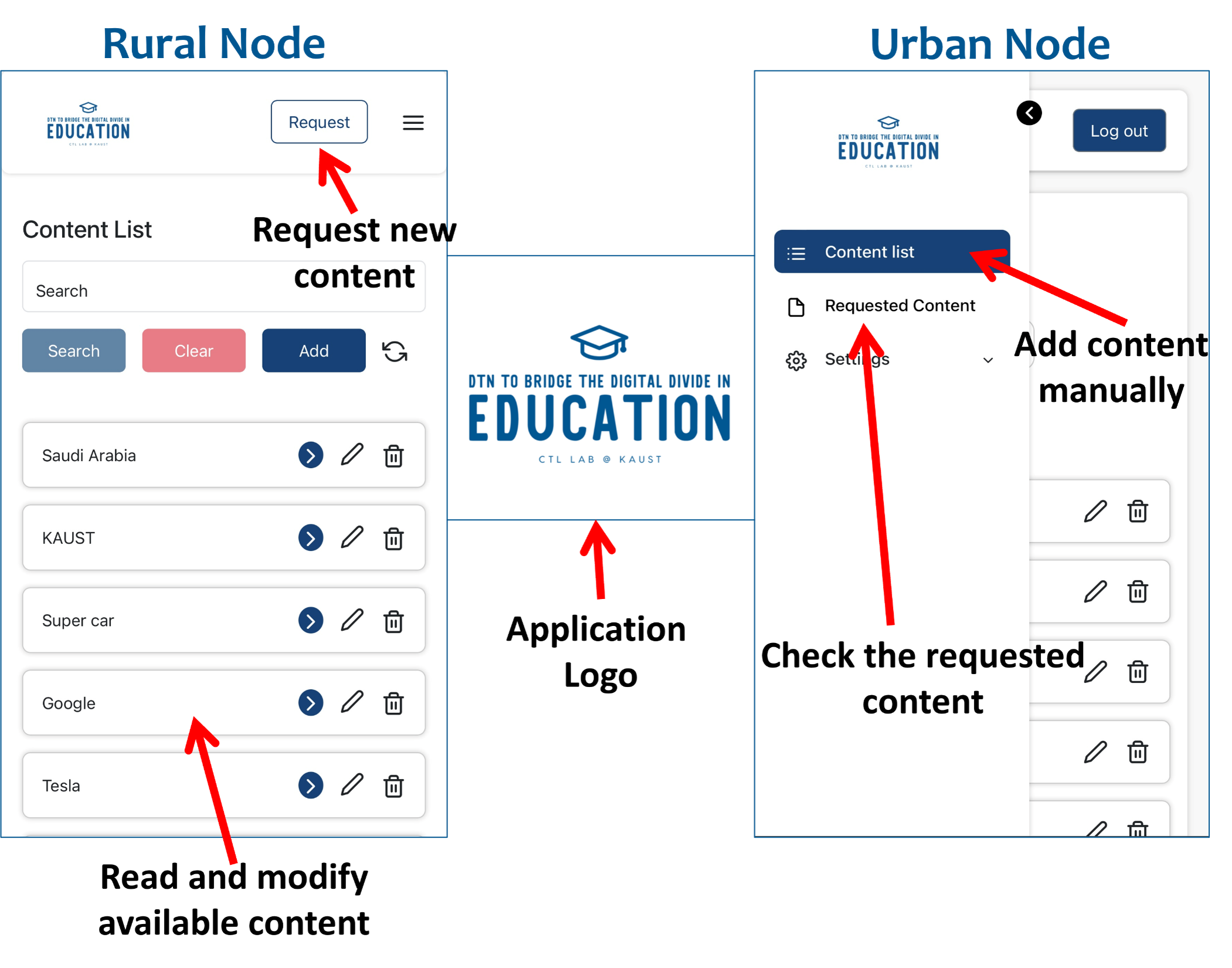}
    \caption{
    The web-based application we designed. For a node in a rural area (Rural Node), the application allows users to create new content, modify previous content, and request new content from the node in the urban area with Internet access (Urban Node). The request will be sent via DTN, carried by the data mules. 
    }
    \label{fig:application_view}
\end{figure}

This section includes an overview of the developed application components on the urban and rural nodes.
These are the software pieces that make up the complete application we developed:
\begin{inparaenum}[i)]
\item Urban node: This is the node located in the urban area with Internet access. We developed a web-based platform that allows the admin to add new topics, modify previous topics, and also automatically fetch Wikipedia requests received from users in rural areas. 
Once the requested topic is fetched from Wikipedia, it will be sent through the DTN network to the rural node. 
Fig. \ref{fig:application_view} shows the urban node interface.
\item Rural node: This is the node located at the schools in rural areas. We developed a web-based application with a list of content and request functionality for the rural node. This content is either added locally by users or created by the urban nodes and delivered to the rural node via DTN. The application also supports requesting newer topics from Wikipedia, where users can type the requested topic. These requests will then be sent via DTN to the urban node, where they will be fetched from Wikipedia through the Internet. The rural node interface of the application is depicted in Fig. \ref{fig:application_view}.
\end{inparaenum}

\subsection{Real-World Testing for Rural Education Solutions}
\begin{figure}[!h]
    \centering
    \includegraphics[width=1\linewidth]{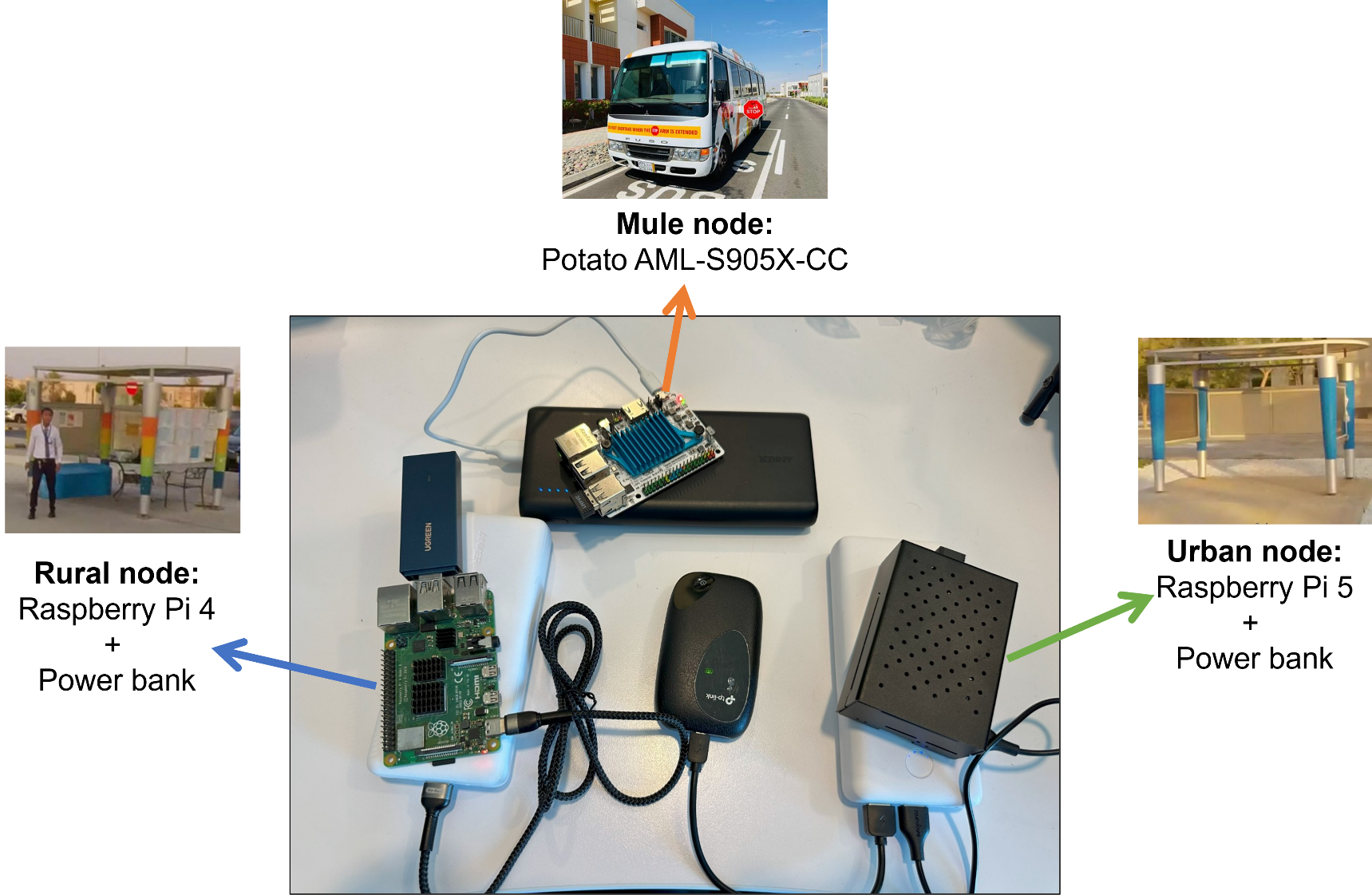}
    \caption{
    The sustainable hardware setup: (1) Urban node manages content, (2) Mule node travels on buses, (3) Rural node provides local access. All components use low-power, affordable hardware suitable for rural deployment.
    }
    \label{fig:application_prototype}
\end{figure}

To better understand how the system works, we designed a prototype consisting of three nodes: the urban node, the mule node, and the rural node. The urban node was hosted on a Raspberry Pi 5, whereas the rural node was hosted on a Raspberry Pi 4. The mule node was implemented on Le Potato AML-S905X-CC. The rural and urban nodes have a Wi-Fi interface to facilitate the connection to the DTN network. In contrast, the urban DTN node has two Wi-Fi interfaces, one to connect to the DTN network and one to connect to the Internet, which is provided on our university campus.
We also had a small Wi-Fi access point connected to both the urban and rural nodes, where users connected to the same local network would have access to the application and act as the DTN facility's common medium. The prototype components are shown in Fig.~\ref{fig:application_prototype}.
\par
\begin{figure}[h!]
    \centering
    \includegraphics[width=0.9\linewidth]{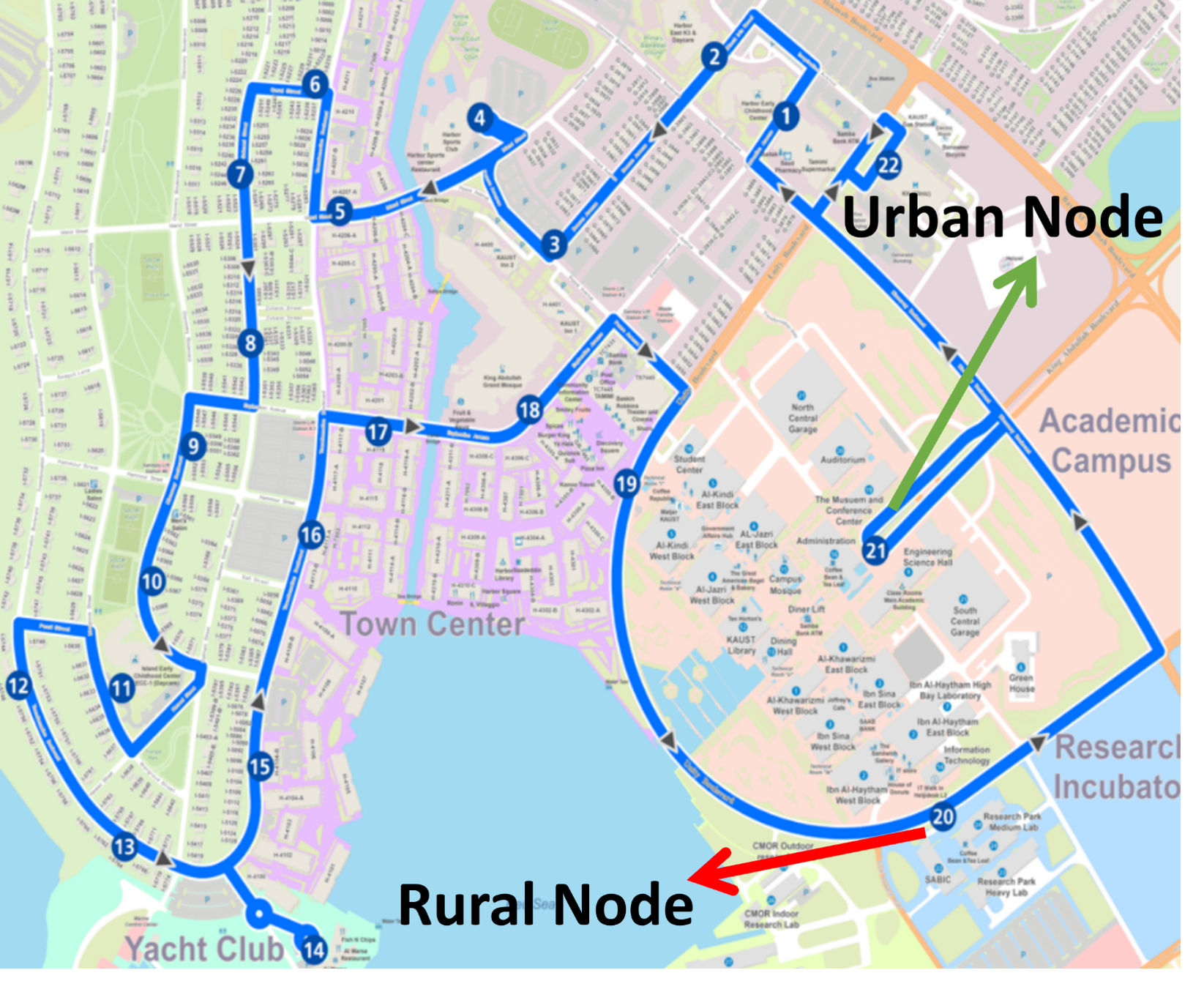}
    \caption{
    Test route showing how data travels between locations. The bus completes a full cycle every 40 minutes, demonstrating how existing transportation can enable digital connectivity.
    } 
    \label{fig:map_prototype}
\end{figure}

To mimic the DTN data mule between urban and rural areas, we placed the urban and rural nodes at two distant bus stops on our university campus, and one node was placed inside one of our campus buses (refer to Fig. \ref{fig:application_prototype}). The DTN network was configured to traverse the data between the urban and rural nodes via the DTN mule node placed in the bus. 
Each node has a power bank to run the Raspberry Pi and the access point routers. 
The bus stops at each bus stop for a few seconds before departing to the next one, and it is scheduled to pass by each of the DTN nodes every 40 minutes. 
Figure \ref{fig:map_prototype} illustrates the locations where we deployed the urban and rural nodes and bus routes on our campus. 

\subsection{Key Lessons for Rural Implementation}
In our experimental validation of the DTN-enabled digital learning platform, we observed that requests for new topics from rural nodes and updates from the urban node were transmitted seamlessly across the network. The implementation, designed as a prototype, provided several insights into the system's performance and scalability.
DTN operate on a store-and-forward mechanism, where data is stored at each node until it can be relayed to the next node. This process is inherently dependent on three key factors: (1) the \textit{contact time}, which refers to the duration of connectivity between nodes; (2) the physical layer connection type, which determines the data rate; and (3) the volume of data (i.e., the size of data stored at each node). In our prototype, moderately priced Wi-Fi interfaces with data rates in the tens of Mbps were employed for the DTN nodes. While this setup facilitated data transfer, it also highlighted the need for extended connection times to ensure complete data transmission.
A significant challenge arose from the stochastic nature of the DTN data mule's operation, which was deployed on a campus bus. The bus's stopping times at each station varied, ranging from a few seconds to up to tens of seconds, depending on the passengers' activity. In some instances, the bus did not stop long enough to allow the DTN nodes to establish a connection and transfer all pending data. This variability in contact time has been extensively analyzed in \cite{DTN_Abdeljabar2025}, which investigates the impact of contact time and data mule frequency on communication performance metrics.
Our initial design prioritized text-based content to minimize network traffic and reduce the reliance on multiple DTN mules. The average data size per educational content ranged between 10 and 30 MB. However, as the platform evolves to incorporate multimedia content (e.g., audio and video) and additional services, the system must be redesigned to handle higher traffic volumes. This will necessitate more robust physical layer connections and higher transmission rates, particularly since the communication window (contact time) between DTN nodes is not controllable. 
Scalability can be achieved by leveraging existing infrastructure, such as deploying additional DTN nodes on more mules traversing the same routes, which would enhance performance metrics, including transmission rates and freshness of information~\cite{DTN_Abdeljabar2025}.

\subsection{Scaling for Greater Impact}
The testing results highlight practical strategies for maximizing the system's societal impact through careful scaling. In the short term, prioritizing text-based educational content ensures dependable performance under existing connectivity limitations. As infrastructure improves, the introduction of additional transport routes or data nodes can facilitate the gradual integration of richer multimedia resources. To enhance the system’s effectiveness, it should be supported by digital literacy programs that equip both students and educators with the skills needed to engage with digital learning platforms.
The study demonstrates that intermittent connectivity, when approached with thoughtful design, can serve as a viable means to bridge the educational divide in underserved communities. A key advantage of the proposed framework lies in its adaptable structure, which enables communities to start with essential learning materials and expand to more advanced content as local conditions evolve. Combined with its low implementation cost and minimal power requirements, the platform presents a sustainable and scalable model for advancing equitable access to education. This suggests that similar frameworks can be adapted to various geographic and cultural contexts, provided they are aligned with the specific needs and priorities of the communities they aim to serve.

\section{Beyond Education: Transforming Rural Communities}
\begin{figure}[h!]
    \centering
    \includegraphics[width=0.8\linewidth]{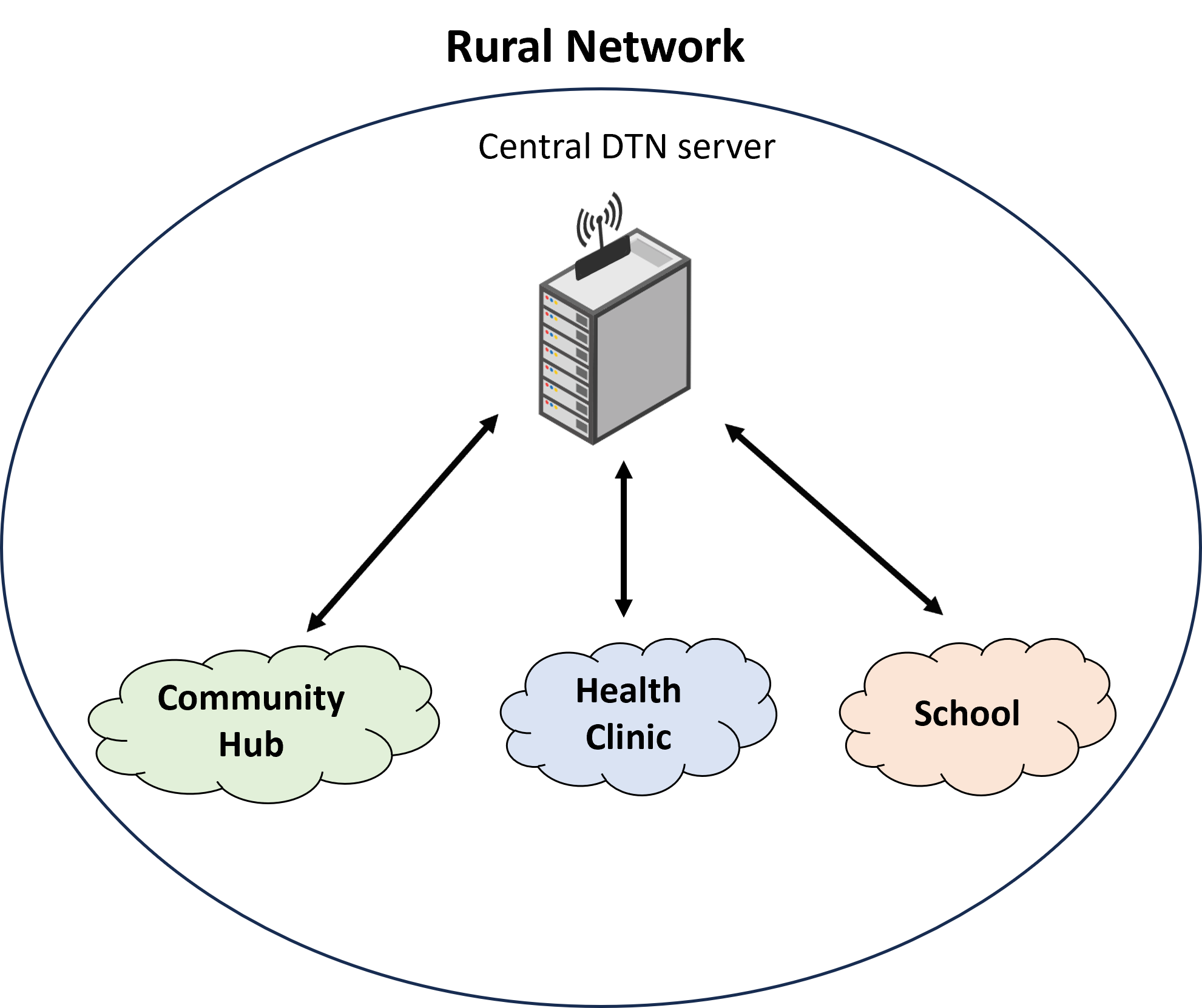}
    \caption{
    A shared connectivity hub enabling equitable digital access. A central DTN node serves schools, clinics, and local businesses, supporting education, healthcare, and economic inclusion through coordinated data transfer.
    }
    \label{fig:DTN_LocalNode}
\end{figure}

While the proposed framework aims to improve access to educational resources in rural areas, it also opens up opportunities for various other services. It empowers rural communities to tap into a range of non-real-time Internet services, enabling them to stay connected and participate in the digital age. 
The DTN facility can allow health clinics to update the required records for patients and get medical advice from remote doctors, and access the latest treatment guidelines. 
Community centres can leverage the same infrastructure to provide valuable services, including offering movie nights using content delivered through the system, and creating important social gathering opportunities in isolated areas. More practically, the framework can distribute agricultural advice, weather forecasts, and market prices to farmers, directly impacting livelihoods and food security.
Small businesses stand to benefit significantly from this framework. The system enables entrepreneurs to create digital catalogues of their products, which can then be shared with urban markets. 
The framework can also support mobile banking services and financial literacy programs, crucial tools for economic empowerment.
To facilitate such an approach to support multiple services, DTN nodes are placed at schools, health clinics and community hubs as illustrated in Fig. \ref{fig:DTN_LocalNode}. 
These nodes are connected to a central DTN node, which can act as the main DTN gateway for each remote village.
This approach can fully capitalize on the DTN infrastructure to support multiple services, ensuring overall growth for the rural communities.

\section{Societal Context and Ethical Imperatives}
The absence of Internet access in rural areas is not only a technological limitation but also a social and ethical issue. According to the United Nations, access to the Internet is essential for achieving universal access to education and ensuring inclusive development \cite{UNDigitalDivide2025}. Lack of connectivity perpetuates educational inequality and undermines the right to information \cite{UNICEFGender2024}.
Digital exclusion reinforces pre-existing inequalities related to income, geography, gender, and disability \cite{sciencedirect2023_internetnonuse, UNDigitalDivide2025}. For instance, in many low-income regions, girls are disproportionately affected by the lack of digital access, which limits their opportunities for educational advancement and professional development \cite{unnews_girls_offline, UNICEFGender2024}. This digital marginalization has intergenerational consequences and directly impedes progress toward achieving global education and equity goals.
In deploying such systems, it is also important to consider ethical challenges related to data security and privacy. While DTN networks are inherently decentralized, they must still ensure that learners’ personal data and educational histories are protected. Designing protocols that respect user confidentiality while enabling useful analytics is key to maintaining trust and equity in digital learning environments.

\section{Conclusion}
This paper demonstrates the potential of a DTN-enabled digital learning platform to enhance educational access in rural areas with intermittent or a lack of Internet connectivity. Our prototype highlighted the effectiveness of DTN as a backhaul-like connectivity solution to deliver educational content to rural areas. 
Our initial prototype focuses on delivering text-based content to rural areas, setting the stage for future enhancements to include multimedia resources, increasing system requirements.
The proposed framework is a step towards bridging the digital divide in education, ultimately leading to sustainable development and prosperity for rural areas.
However, realizing these benefits requires careful attention to the social and sustainability dimensions of technological innovation.
Moreover, partnerships with public education authorities, non-governmental organizations, and community groups can guide policy frameworks that support the responsible scaling of such infrastructure. Policies that recognize intermittent connectivity as a viable access model can help accelerate deployment in areas where continuous Internet access remains a long-term challenge.

\ifCLASSOPTIONcaptionsoff
  \newpage
\fi

\bibliographystyle{IEEEtran}
\bibliography{references}

\begin{IEEEbiography}[{\includegraphics[width=1in,height=1.25in,clip,keepaspectratio]{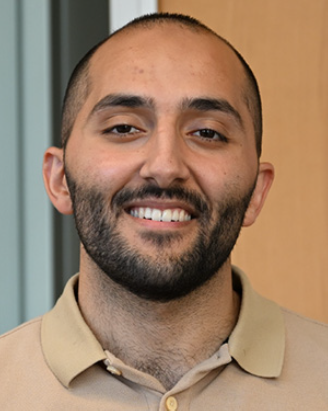}}]{Salah Abdeljabar}
(Graduate~Student~Member, IEEE) received the B.Sc. degree in electrical engineering from The University of Jordan, Amman, Jordan, in 2019, and the M.Sc. degree in electrical and computer engineering from the King Abdullah University of Science and Technology, Thuwal, Saudi Arabia, in 2023, where he is currently pursuing the Ph.D. degree. His research interests include long-range (LoRa) communication, Delay delay-tolerant networking (DTN), and optical wireless communications systems.
\end{IEEEbiography}

\begin{IEEEbiography}[{\includegraphics[width=1in,height=1.25in,clip,keepaspectratio]{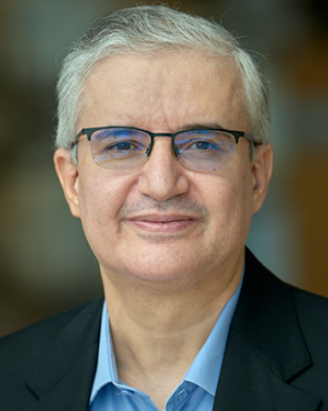}}]{Mohamed-Slim Alouini}
(Fellow, IEEE) was born in Tunis, Tunisia. He received the Ph.D. degree in electrical engineering from the California Institute of Technology, Pasadena, CA, USA, in 1998. He served as a Faculty Member with the University of Minnesota, Minneapolis, MN, USA, then with Texas A\&M University at Qatar, Doha, Qatar, before joining the King Abdullah University of Science and Technology, Thuwal, Makkah, Saudi Arabia, as a Professor of Electrical Engineering in 2009. His current research interests include the modeling, design, and performance analysis of wireless communication systems.
\end{IEEEbiography}

\end{document}